\documentstyle[aps,prl,psfig,multicol]{revtex}
\begin{document}
\newenvironment{tab}[1]
{\begin{tabular}{|#1|}\hline}
{\hline\end{tabular}}

\title {Interface effects in superconductor-nanotubes hybrid structures}

\author{N. Stefanakis}
\address{e-mail:stefan@iesl.forth.gr}
\date{\today}
\maketitle

\begin{abstract}
The objective of the present paper 
is to investigate the proximity effect in junctions
of superconductors with carbon nanotubes. The method is the 
lattice BdG equations within the Hubbard model. The proximity 
effect depends sensitively on the connection giving the  
possibility to control the proximity effect by performing simple 
geometrical changes in the hybrid system. 
\end{abstract}

\section{Introduction}
In the proximity effect the superconducting pair amplitude
appears in a region where the pair interaction is zero.
During the last decades it has been investigated in several
mesoscopic structures e.g. metallic wires made of normal metals
between two macroscopic superconducting electrodes \cite{courtois}.
More recently it was probed in superconductor ferromagnet
hybrid structures where the pair amplitude shows decaying oscillations
with alternating sign inside the ferromagnetic layer \cite{kontos}.

In the past decade transport measurements were used to explore the
properties of nanometer-scale structures. An example is the carbon
nanotube \cite{iijima}. A carbon nanotube is obtained from a slice of
graphene sheet wrapped into a seamless cylinder. The conducting properties
depend sensitively on the diameter and the helicity. They can be classified
to 'armchair', 'zigzag', and 'chiral' depending on their
wrapping vector. They come in two forms,
the multiwall (MWNT) with diameter $10-30$nm and the single wall
(SWNT) with diameter $1-2$nm. Due to their small diameter SWNT provide
ideal systems to study transport properties of 1D conductors.

Is is possible to create superconducting junctions
with SWNT embedded between superconducting contacts.
In superconductor - SWNT - superconductor junctions, proximity
induced superconductivity has been observed \cite{kasumov}. 
The temperature and
magnetic field dependence of the critical current shows
unusual features due to their strong one dimensional character.
More explicitly the critical current exceeds the predicted one for
SNS junctions by a large factor.
Also the temperature and magnetic field dependence
of the critical current is almost linear.
In contrast in superconductor - SWNT - superconductor junctions
with high transparent interfaces a dip in a 
differential resistance was observed \cite{morpurgo}. 
This was attributed to Andreev reflection processes.
When the transparency was low a peak was observed due to 
normal tunneling processes. 
Recently observation of 1D superconductivity in single-walled 
$4 A$ carbon nanotubes was reported \cite{tang}.

In this paper we describe the proximity effect
in superconductor - SWNT
junctions by solving the numerically Bogoliubov de Gennes equations within
the Hubbard model. We calculate the local density of states (LDOS)
and the pair amplitude self consistently. We find that 
the LDOS is strongly modified in the interface of 
superconductor with the nanotube and depends sensitively on the 
pairing symmetry of the superconductor, the chirality of the nanotube, 
and the connection between the two structures giving the opportunity 
to control the electronic properties at nanometer scale 
by performing topological changes to the hybrid system. The junctions 
of superconductor with nanotubes differ from the conventional 
junctions in their cross section which is of nanometer scale.

In the following we describe the numerical method in Sec II.
In Sec III we present the results for the proximity 
effect in 
superconductor-side(-end) connected nanotube 
junction, we examine the effect of the pairing symmetry 
and chirality
and finish with the conclusions.

\section{method}
In our paper we describe the proximity effect
in superconductor - SWNT
junctions by solving the Bogoliubov de Gennes equations within
the Hubbard model \cite{stefan,stefan1,stefan2}.
The Hamiltonian for the extended Hubbard model on a two dimensional square
lattice is
\begin{eqnarray}
H & = & -t\sum_{<i,j>\sigma}c_{i\sigma}^{\dagger}c_{j\sigma}
+\mu \sum_{i\sigma} n_{i\sigma} \nonumber \\
  & + & V_0\sum_{i} n_{i\uparrow} n_{i\downarrow}
+\frac{V_1}{2}\sum_{<ij>\sigma\sigma^{'}} n_{i\sigma} n_{j\sigma^{'}}\,
,~~~\label{bdgH}
\end{eqnarray}
where $i,j$ are sites indices and the angle brackets indicate that the
hopping is only to nearest neighbors,
$n_{i\sigma}=c_{i\sigma}^{\dagger}c_{i\sigma}$ is the electron number
operator in site $i$, $\mu$ is the chemical potential, and $V_0$, $V_1$
are on site and nearest-neighbor interaction strength. Negative values
of $V_0$ and $V_1$ mean attractive interaction and positive values mean
repulsive interaction.
Within the mean field approximation Eq. (\ref{bdgH}) is reduced  to
the Bogoliubov deGennes equations:
\begin{equation}
\left(
\begin{array}{ll}
  \hat{\xi} & \hat{\Delta} \\
  \hat{\Delta}^{\ast} & -\hat{\xi}
\end{array}
\right)
\left(
\begin{array}{ll}
  u_n(r_i) \\
  v_n(r_i)
\end{array}
\right)
=\epsilon_n
\left(
\begin{array}{ll}
  u_n(r_i) \\
  v_n(r_i)
\end{array}
\right)
,~~~\label{bdgbdg}
\end{equation}
such that
\begin{equation}
\hat{\xi}u_n(r_i)=-t\sum_{\hat{\delta}}
u_n(r_i+\hat{\delta})+\mu u_n(r_i),~~~\label{bdgxi}
\end{equation}

\begin{equation}
\hat{\Delta}u_n(r_i)=\Delta_0(r_i)u_n(r_i)+\sum_{\hat{\delta}}
\Delta_{\delta}(r_i)u_n(r_i+\hat{\delta}),~~~\label{bdgdelta}
\end{equation}
where the gap functions are defined by

\begin{equation}
\Delta_0(r_i)\equiv
V_0<c_{\uparrow}(r_i)c_{\downarrow}(r_i)>,~~~\label{bdgdelta0}
\end{equation}
\begin{equation}
\Delta_{\delta}(r_i)\equiv
V_1<c_{\uparrow}(r_i+\hat{\delta})c_{\downarrow}(r_i)>,~~~\label{bdgdeltadelta}
\end{equation}
where $\hat{\delta}=\hat{x},-\hat{x},\hat{y},-\hat{y}$. Equations
(\ref{bdgbdg}) are subject to the self consistency requirements
\begin{equation}
\Delta_0(r_i)=V_0\sum_{n}
u_n(r_i)v_n^{\ast}(r_i)\tanh\left(\frac{\beta
\epsilon_n}{2}\right),~~~\label{bdgselfD0}
\end{equation}

\begin{eqnarray}
\Delta_{\delta}(r_i) & = & \frac{V_1}{2}\sum_{n}
(u_n(r_i+\hat{\delta})v_n^{\ast}(r_i) \nonumber \\
 & + & u_n(r_i)v_n^{\ast}(r_i+\hat{\delta}) )\tanh\left(\frac{\beta
\epsilon_n}{2}\right) .~~~\label{bdgselfDdelta}
\end{eqnarray}
We start with the approximate initial conditions for the gap functions
(\ref{bdgselfD0},\ref{bdgselfDdelta}). After exact diagonalazation of Eq.
(\ref{bdgbdg})
we obtain the $u(r_i)$ and
$v(r_i)$ and the eigenenergies $\epsilon_n$.
The quasiparticle amplitudes are then inserted into Eq.
(\ref{bdgselfD0},\ref{bdgselfDdelta}) and new
gap functions $\Delta_0(r_i)$ and $\Delta_{\delta}(r_i)$ are evaluated.
We reinsert these quantities into Eqs. (\ref{bdgxi},\ref{bdgdelta}),
and we proceed in the same way until we achieve self-consistency:
i.e. when the norm of the difference of $\Delta_0(r_i)$ and
$\Delta_{\delta}(r_i)$ from their previous values 
is less than the desired accuracy.
We then compute the $d$-wave gap
function given by the expression:
\begin{equation}
\Delta_d(r_i)=\frac{1}{4}[\Delta_{\hat{x}}(r_i)+\Delta_{-\hat{x}}(r_i)
-\Delta_{\hat{y}}(r_i)-\Delta_{-\hat{y}}(r_i)],~~~\label{bdgdeltad}
\end{equation}
and the LDOS at the $i$th site which is given by
\begin{equation}
\rho_i(E)=-2\sum_{n}
\left [ |u_n(r_i)|^2 f^{'}(E-\epsilon_n)
+ |v_n(r_i)|^2 f^{'}(E+\epsilon_n) \right ]
,~~~\label{bdgdos}
\end{equation}
where the factor $2$ comes from the twofold spin degeneracy,
$f^{'}$ is the derivative of the Fermi function,
\begin{equation}
f(\epsilon)=\frac{1}{\exp(\epsilon/k_B T) + 1}
\end{equation}.

The SWNT is described as a single sheet of graphite composed of carbon
atoms arranged on the sites of a honeycomb lattice.
Within the tight binding method one orbital is associated
per carbon atom, and a tunneling element $t$ between neighboring
atoms. SWNT are formed by rolling the honeycomb sheet into a
cylinder. We describe armchair or zigzag structures.
The coupling between the two structures is through
single bond or multiple bonds connecting the edge or side sites of the tube
to the 2d superconductor.
We calculate the LDOS, and pair amplitude.

\section{results}
\subsection{1d-superconductivity in SWNT}
We would like to describe the LDOS for a SWNT which exhibits 
superconductivity. Within lattice Hubbard model the presence of 
on site attractive interaction give rise to $s$-wave superconductivity. 
The main characteristic which is visible 
in the LDOS is the presence of gap (see Fig. \ref{LDOS55supra.fig}). 
For the bulk LDOS the gap coexists with bands showing
one-dimensional Van-Hove singularities at the band edges. 
However close to the interface 
the LDOS is modified due to boundary effects.

\subsection{rotation of nanotube with respect to the superconductor}
\begin{figure}
\begin{center}
\leavevmode
\psfig{figure=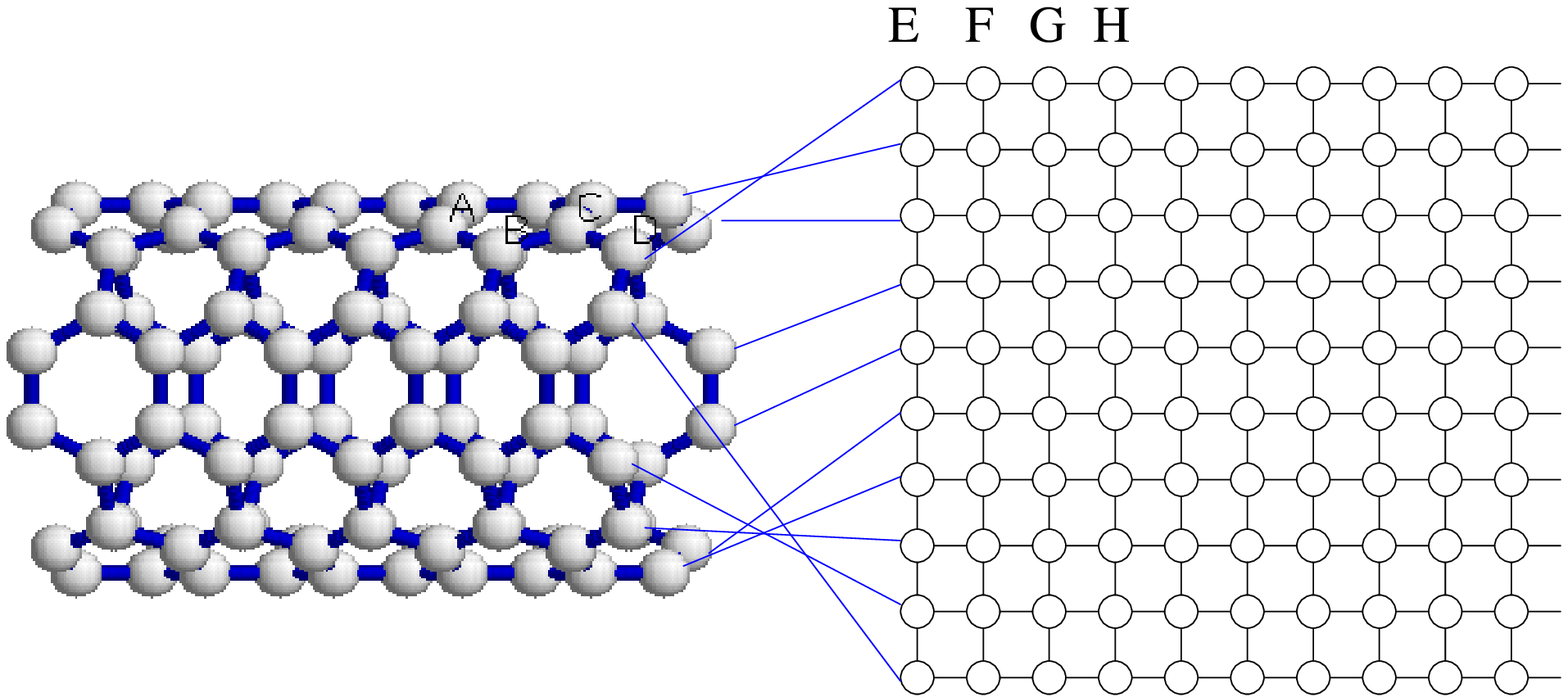,width=10cm,angle=0}
\end{center}
\caption{The
open armchair (5,5)
nanotube composed of 10 layers 
end-connected to a two dimensional s-wave superconductor of 
$10 \times 10$ sites.
}
\label{supra_swnt.fig}
\end{figure}

\begin{figure}
\begin{center}
\leavevmode
\psfig{figure=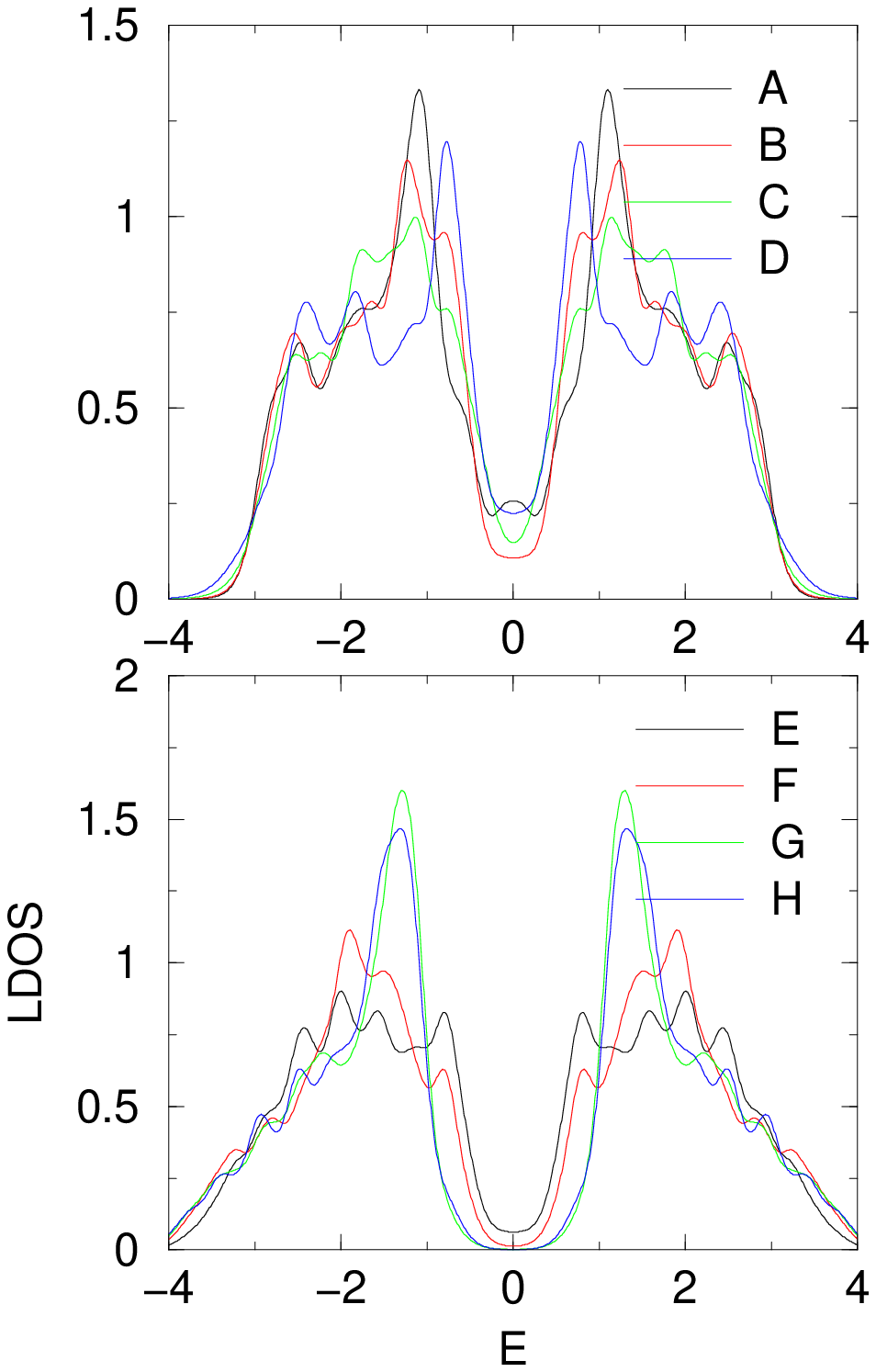,width=6cm,angle=0}
\end{center}
\caption{a) The LDOS for 
the hybrid structure shown in Fig. \ref{supra_swnt.fig}.
The points A,B,C,D belong to the end-connected nanotube
while the points E,F,G,H to the superconductor.
}
\label{LDOS55supra_swnt.fig}
\end{figure}

\begin{figure}
\begin{center}
\leavevmode
\psfig{figure=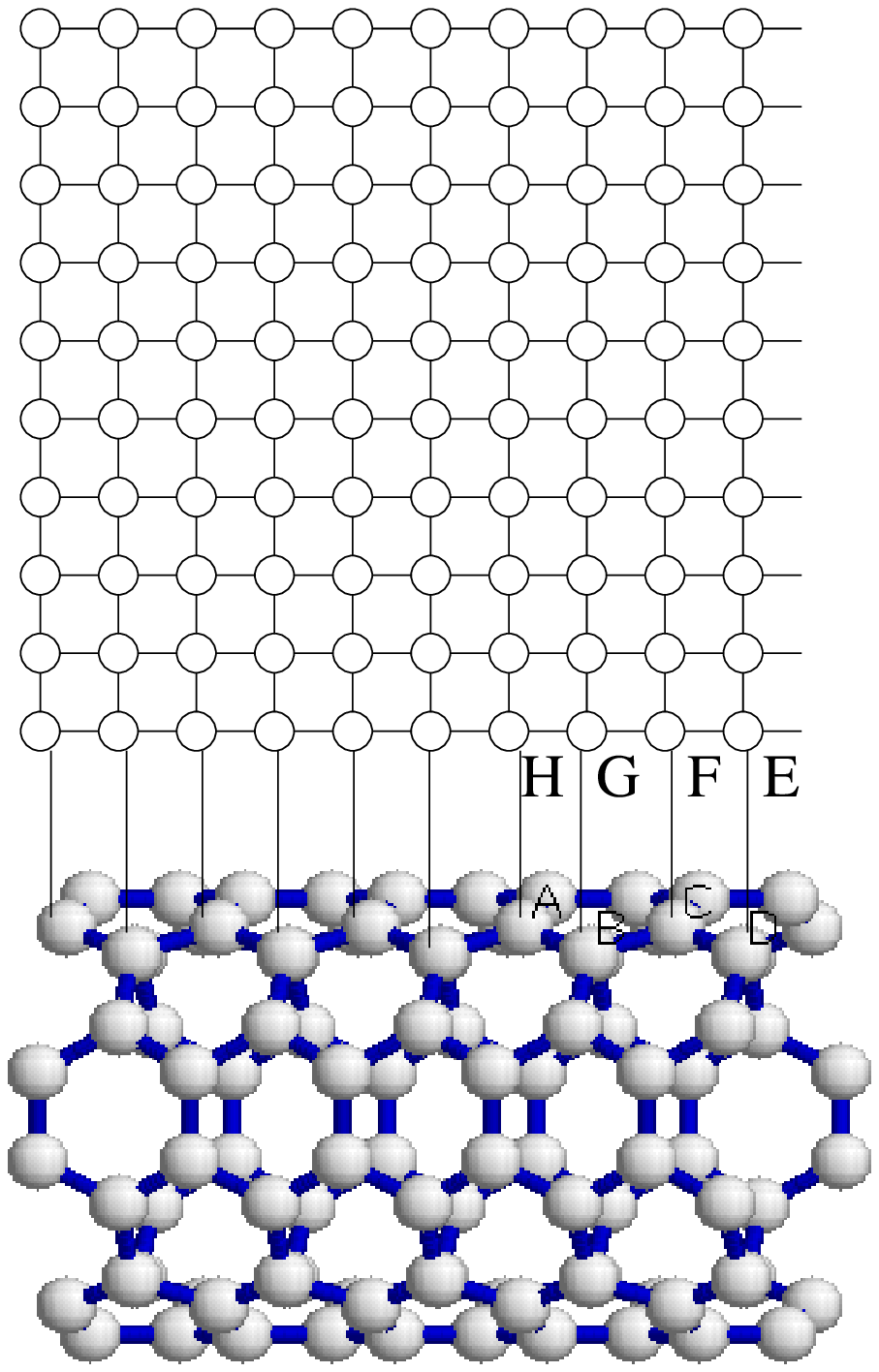,width=6cm,angle=0}
\end{center}
\caption{The open
armchair (5,5) nanotube composed of 10 layers 
side-connected to a two dimensional $10 \times 10$ s-wave superconductor.
}
\label{sidesupra_swnt.fig}
\end{figure}

\begin{figure}
\begin{center}
\leavevmode
\psfig{figure=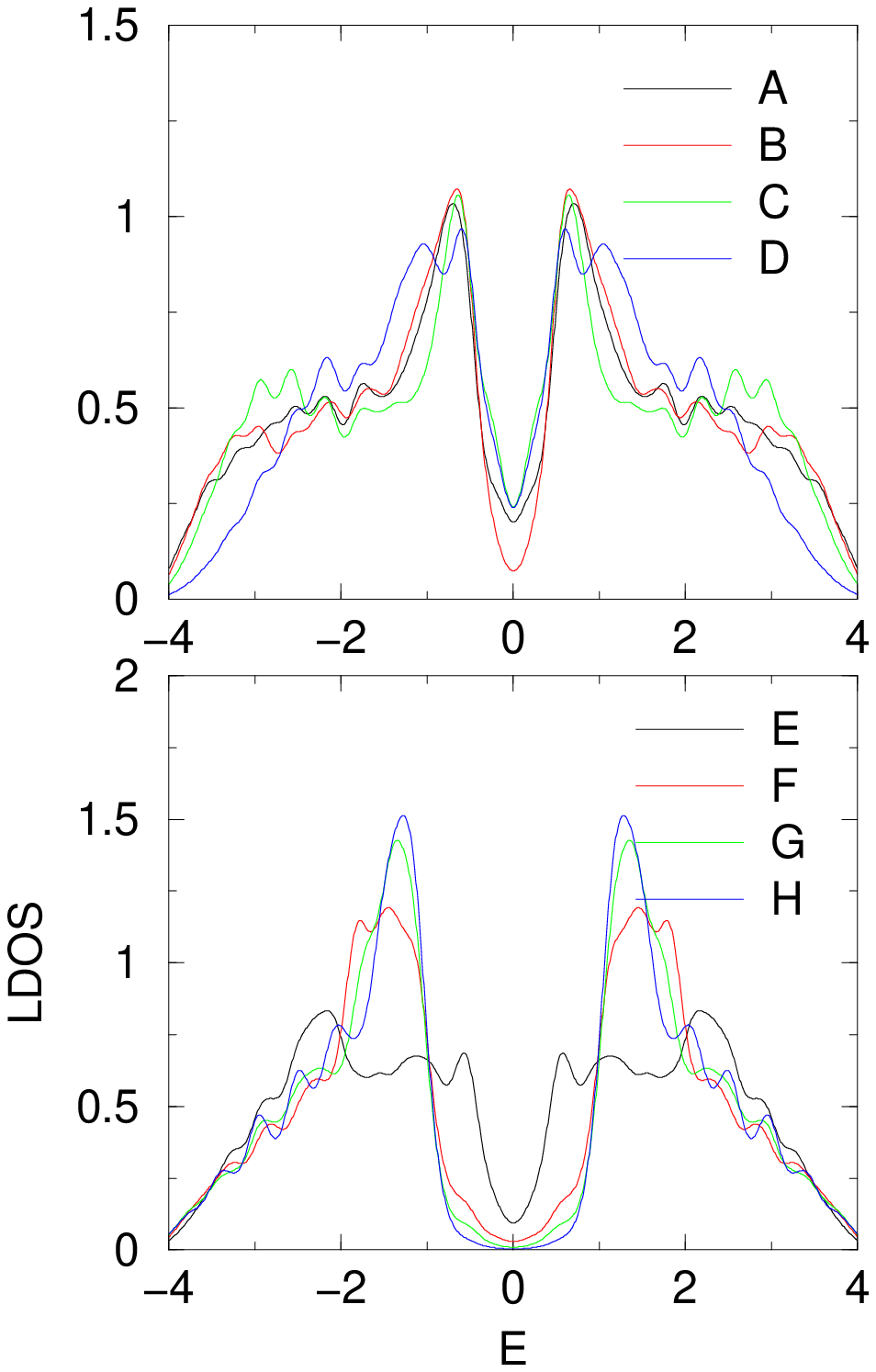,width=6cm,angle=0}
\end{center}
\caption{
The LDOS for the hybrid structure shown in Fig. \ref{sidesupra_swnt.fig}.
The points A,B,C,D belong to the side-connected nanotube
while the points E,F,G,H to the superconductor. 
}
\label{sideLDOS55supra_swnt.fig}
\end{figure}

\begin{figure}
\begin{center}
\leavevmode
\psfig{figure=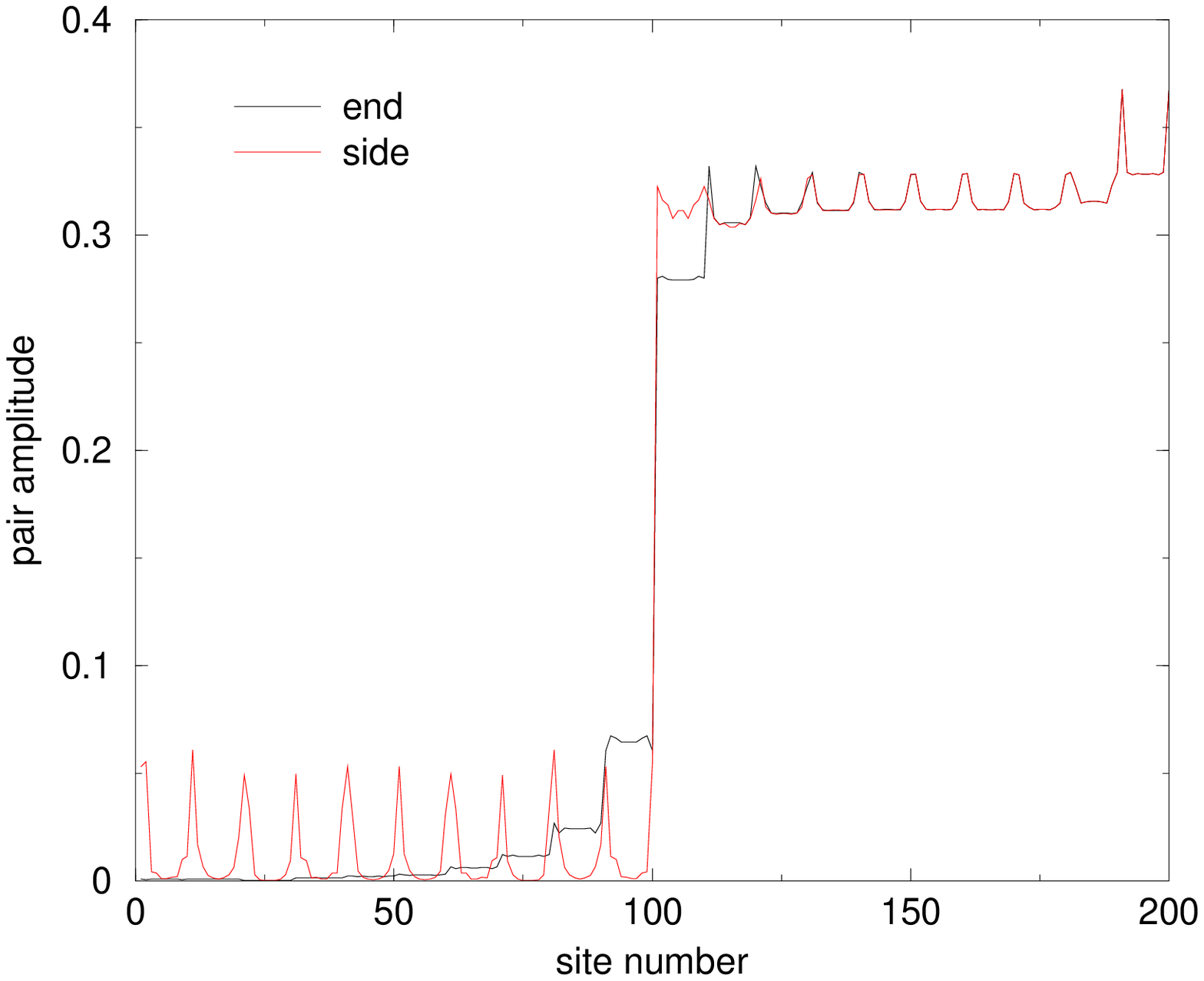,width=6cm,angle=0}
\end{center}
\caption{a) The comparison of the pair amplitude for the 
superconductor side- (end-) connected SWNT.
}
\label{pa55supra_swnt.fig}
\end{figure}

The next step is to describe the proximity effect in superconductor 
SWNT. Here the SWNT shows superconductivity which is due to the 
proximity with a metal that exhibits superconductivity.
We show 
that the LDOS due to proximity between the two structures 
can be modulated by simple geometrical transformations like 
rotation of the SWNT with respect to the superconductor. 
We study first the case of an end-connected SWNT to 
a superconductor as seen in Fig. \ref{supra_swnt.fig}.
We see in Fig. \ref{LDOS55supra_swnt.fig} the crossover from the 
metallic behavior which appears as finite LDOS at zero energy 
to the superconducting state where a gap appears.
The deviation from the metallic behavior which appears as 
finite LDOS at Fermi energy becomes weaker as we go to the bulk.
Next we present the side-connected SWNT 
to a superconductor as seen in Fig. \ref{sidesupra_swnt.fig}.
In the LDOS in Fig. \ref{sideLDOS55supra_swnt.fig} we see that 
it deviates from the metallic behavior and is modulated by the 
distance from the surface. The proximity induced gap in the 
superconductor is of smaller magnitude compared to the 
end-connection case. 
In Fig. \ref{pa55supra_swnt.fig} we compare the pair amplitude for 
the side- and end- connection cases. In the end connection the pair amplitude 
decays toward the bulk of the SWNT showing plateaus across the nanotube. 
In the side 
connection the pair amplitude is almost homogeneous along the nanotube, 
but is varied across the nanotube. 

\begin{figure}
\begin{center}
\leavevmode
\psfig{figure=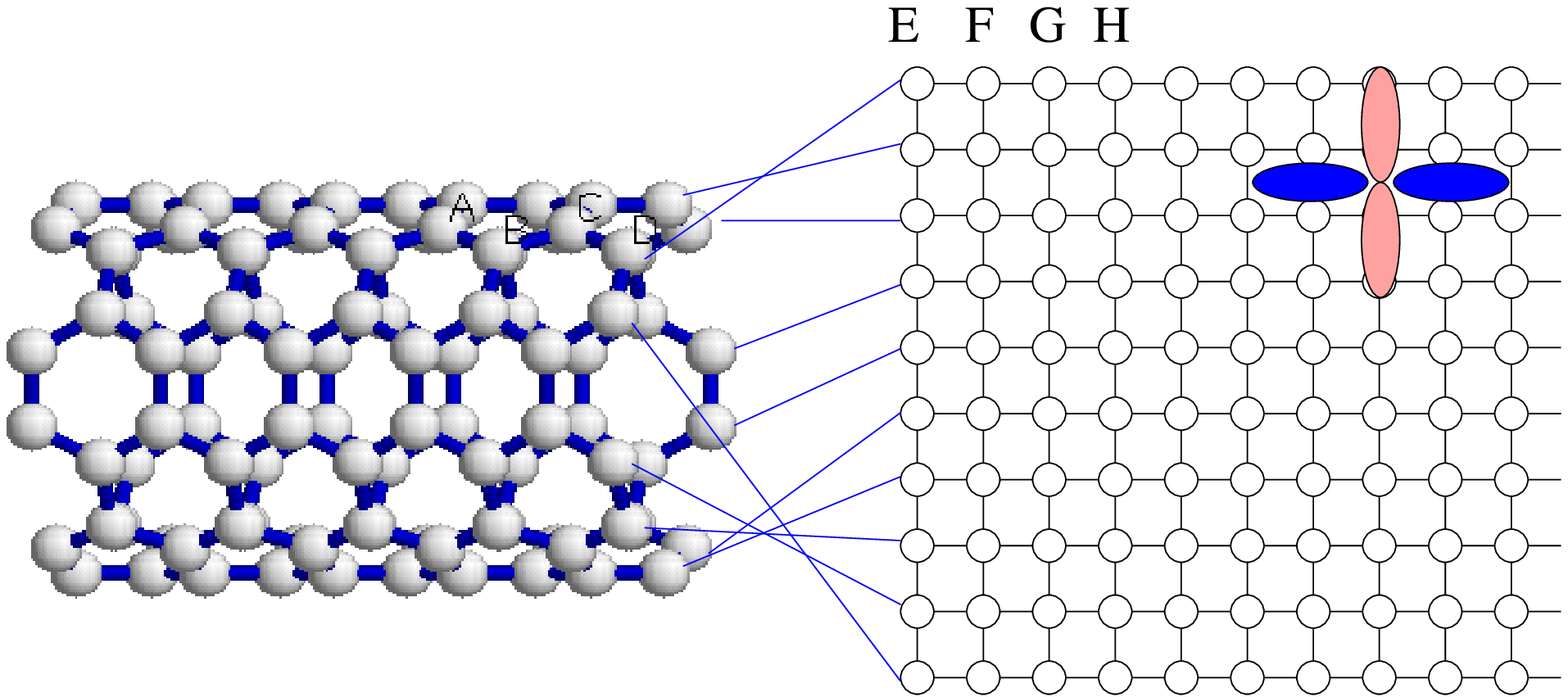,width=10cm,angle=0}
\end{center}
\caption{The
open armchair (5,5) nanotube composed of 10 layers 
end-connected to a two dimensional $10\times 10$ d-wave superconductor.
}
\label{dsupra_swnt.fig}
\end{figure}

\begin{figure}
\begin{center}
\leavevmode
\psfig{figure=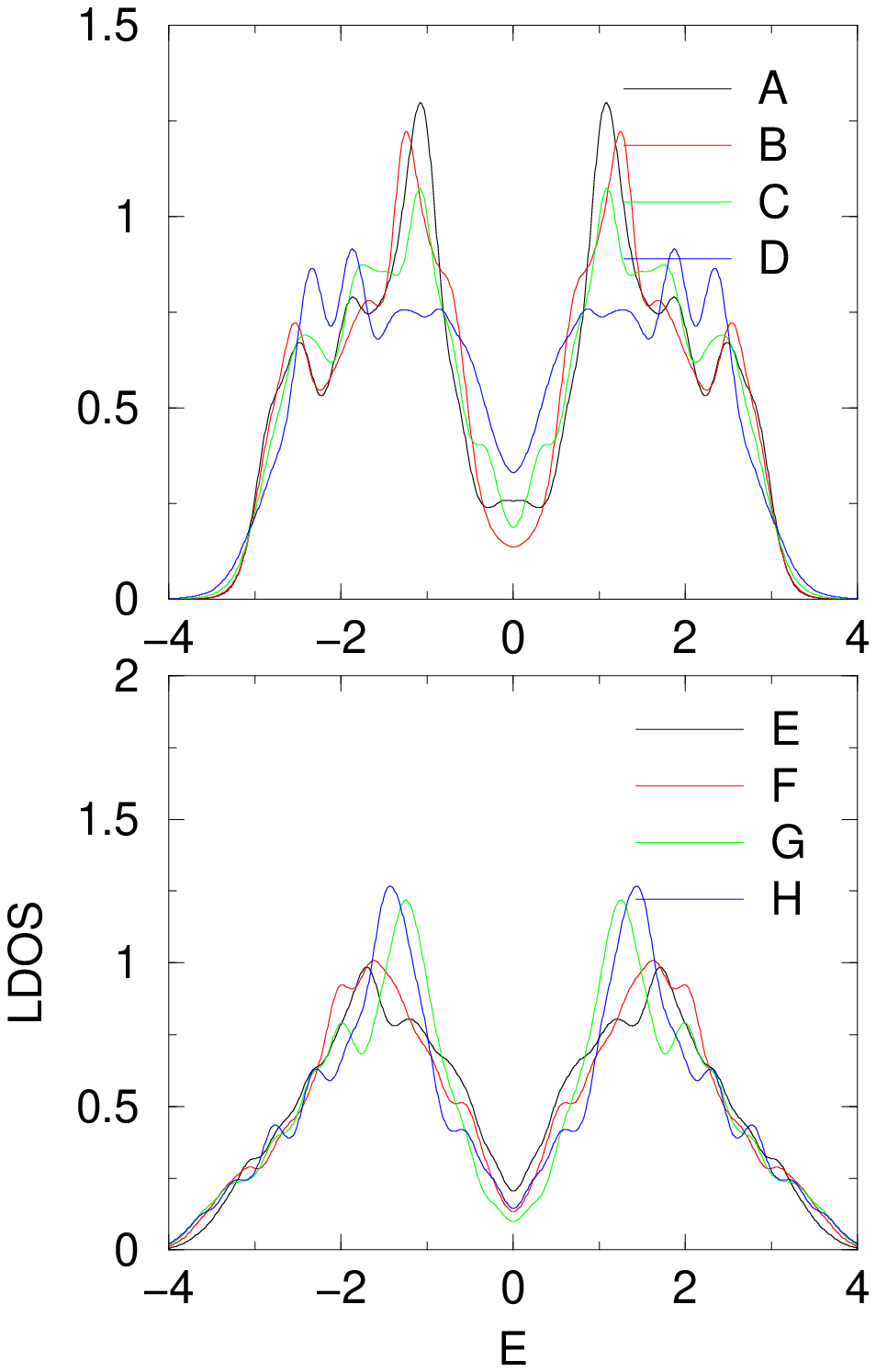,width=6cm,angle=0}
\end{center}
\caption{
The LDOS for the hybrid structure shown in Fig. \ref{dsupra_swnt.fig}.
The points A,B,C,D belong to the end-connected nanotube
while the points E,F,G,H to the d-wave superconductor.
}
\label{LDOS55dsupra_swnt.fig}
\end{figure}

\begin{figure}
\begin{center}
\leavevmode
\psfig{figure=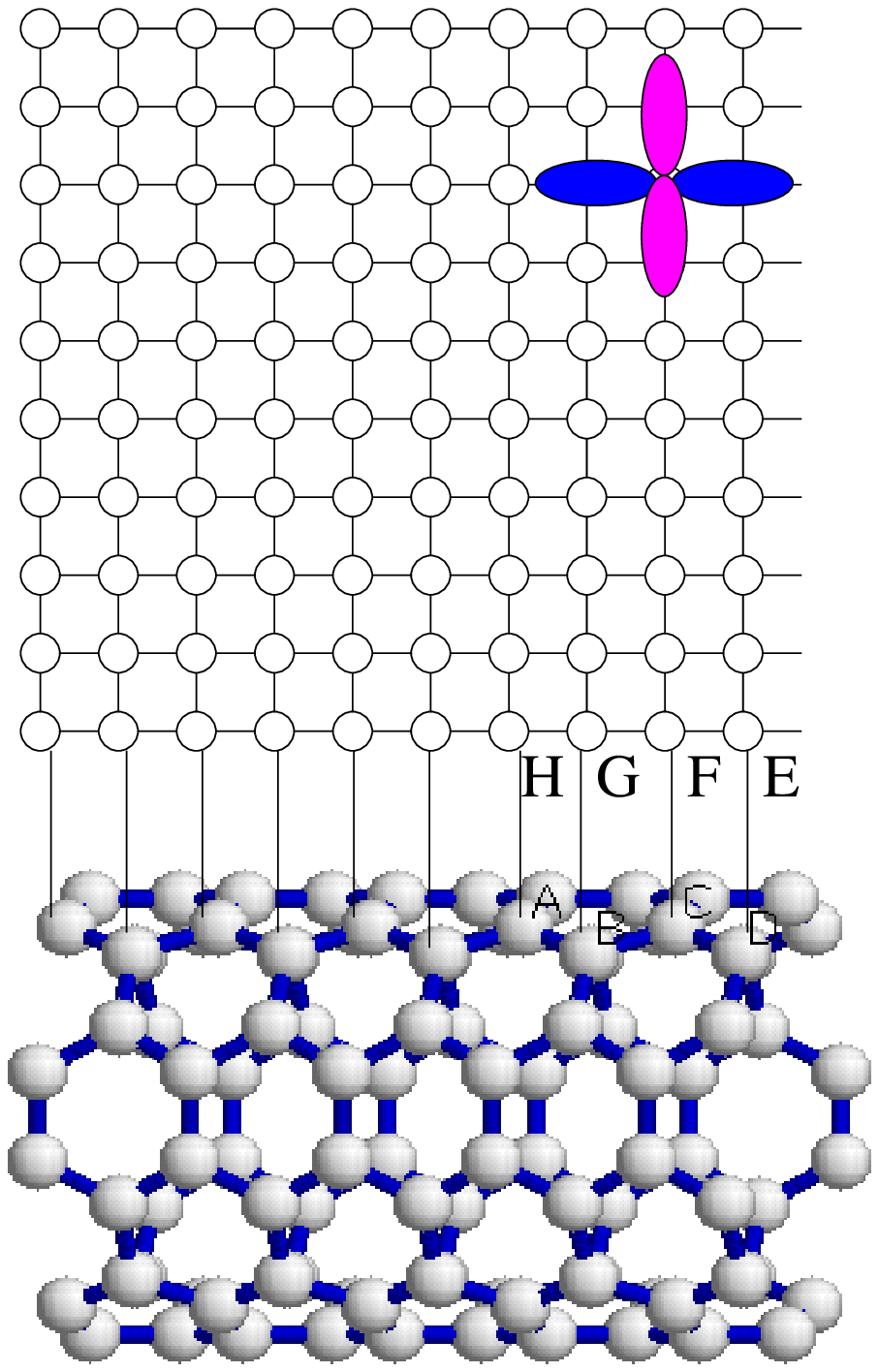,width=6cm,angle=0}
\end{center}
\caption{The open
armchair (5,5) nanotube composed of 10 layers 
side-connected to a two dimensional $10\times10$ d-wave superconductor.
}
\label{sidedsupra_swnt.fig}
\end{figure}

\begin{figure}
\begin{center}
\leavevmode
\psfig{figure=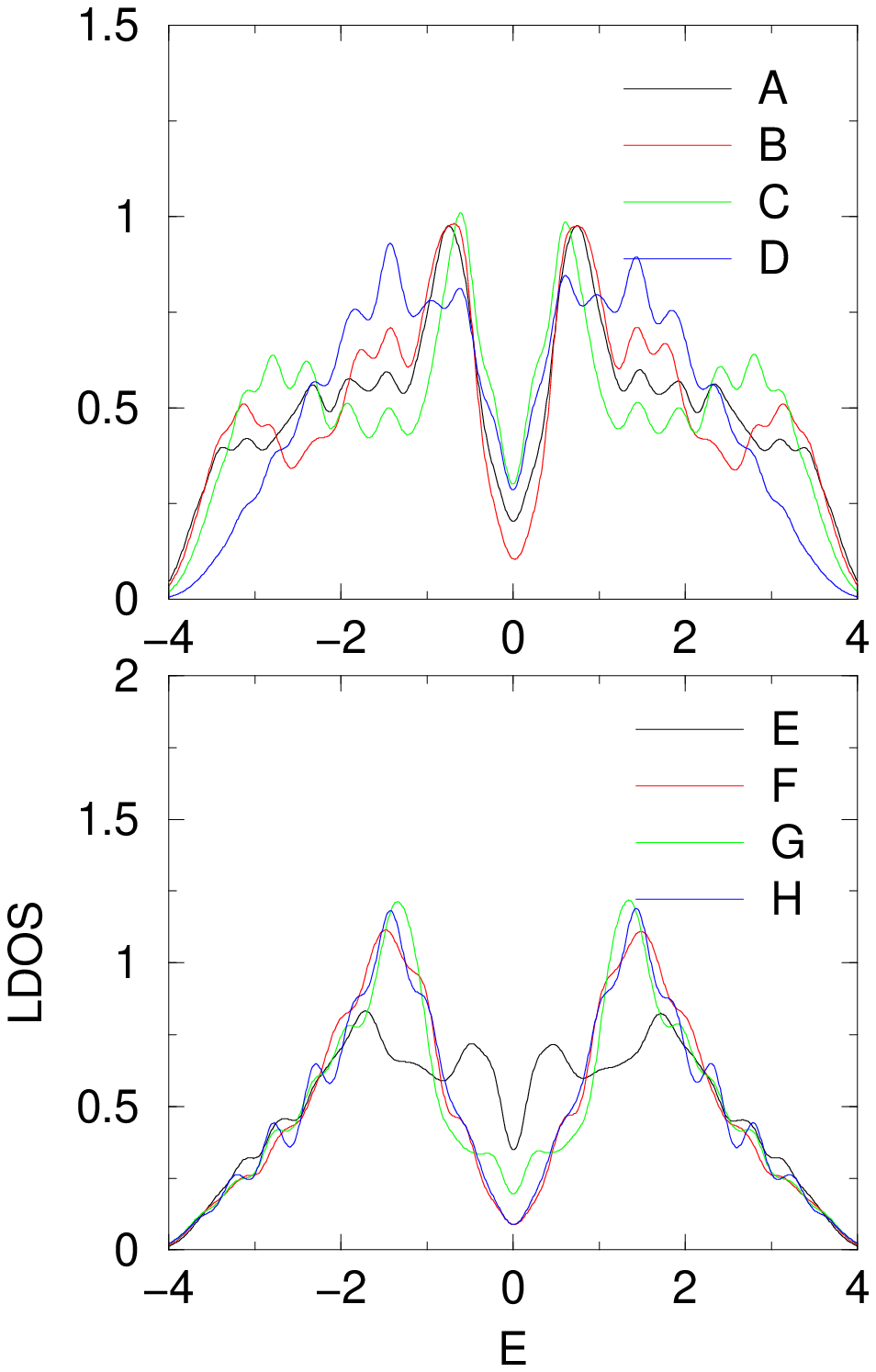,width=6cm,angle=0}
\end{center}
\caption{
The LDOS for the hybrid structure shown in Fig. \ref{sidedsupra_swnt.fig}.
The points A,B,C,D belong to the side-connected nanotube
while the points E,F,G,H to the d-wave superconductor.
}
\label{sideLDOS55dsupra_swnt.fig}
\end{figure}

\begin{figure}
\begin{center}
\leavevmode
\psfig{figure=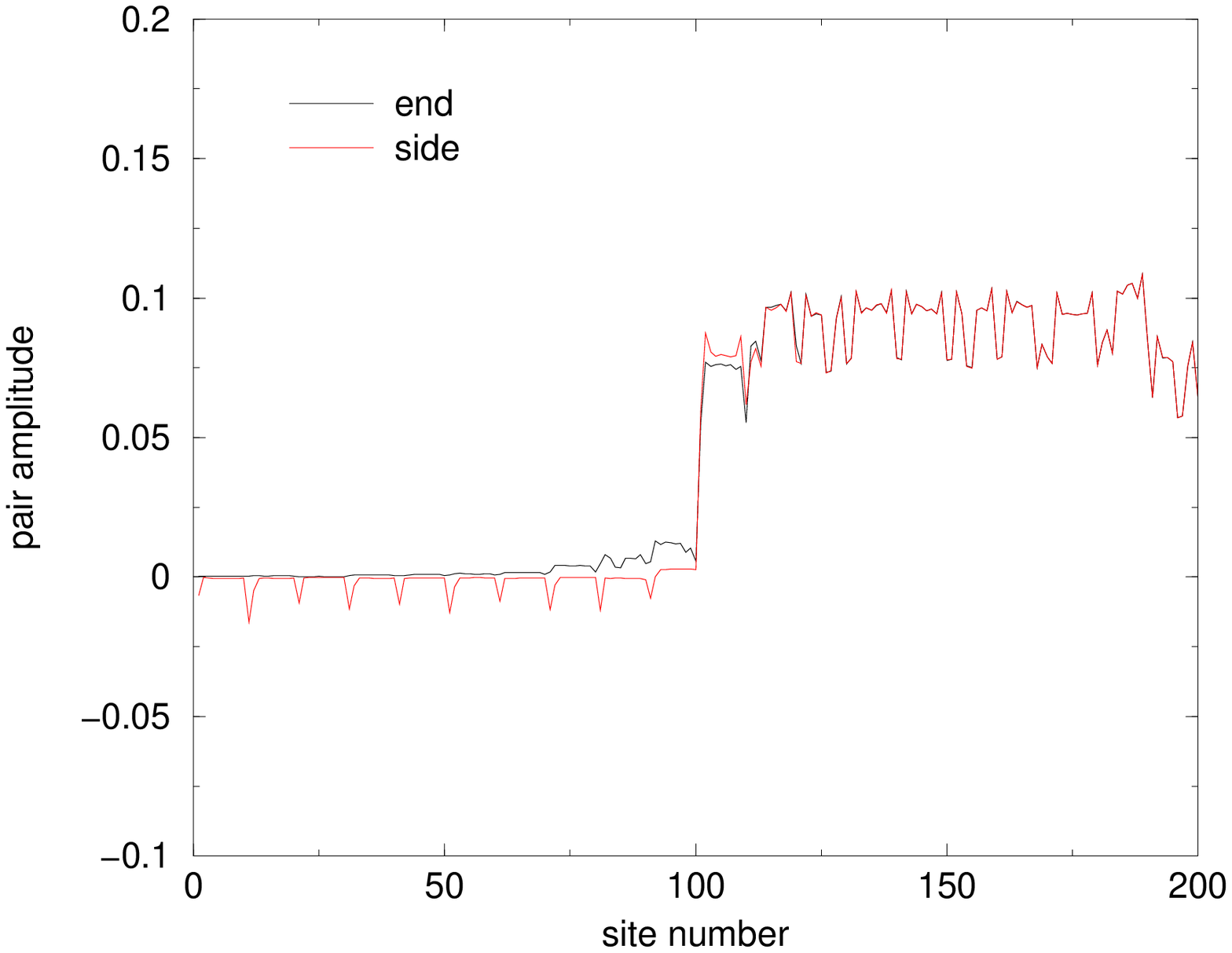,width=6cm,angle=0}
\end{center}
\caption{a) The comparison of the pair amplitude for the 
d-wave superconductor - side- (end-) connected SWNT.
}
\label{pa55dsupra_swnt.fig}
\end{figure}

We see in Fig. \ref{dsupra_swnt.fig} an end connected nanotube to a 
$d$-wave superconductor.
We see in Fig. \ref{LDOS55dsupra_swnt.fig}
that the LDOS 
changes from metallic A to superconducting D like 
where a gap appears as we approach the interface. 
The form of the gap in the LDOS is V like due to the 
presence of nodes in the pair amplitude along certain 
directions in k-space.
Next we present the side-connected SWNT 
to a d-wave superconductor as seen in Fig. \ref{sidedsupra_swnt.fig}.
In the LDOS in Fig. \ref{sideLDOS55dsupra_swnt.fig} we see that 
it deviates from the metallic behavior and is modulated by the 
distance from the surface. Moreover the proximity induced gap 
in the superconductor is of smaller magnitude compared to 
the armchair case. 
In Fig. \ref{pa55dsupra_swnt.fig} we compare the pair amplitude for 
the side- and end- connection cases. In the end connection the pair amplitude 
decays toward the bulk of the SWNT showing plateaus across the nanotube. 
In the side 
connection the pair amplitude is almost homogeneous along the nanotube, 
but is varied across the nanotube. Negative pair amplitude appears 
in the proximity induced superconductivity due to the absence of 
hopping elements in the x direction. 
Therefore 
due to the fact that the pair interaction in d-wave is strongly 
non local and depends on the number of nearest neighbors that 
are available, the induced proximity pair amplitude inside the 
honeycomb lattice is modified for d-wave compared to s-wave.
Concluding this section we could say 
that independently on the pairing symmetry the proximity induced gap 
is smaller for the side connection than the end connection.
The induced pair amplitude is also different for the side than the 
end connection and shows additional features due to the pairing symmetry.

\subsection{effect of chirality of nanotube}
\begin{figure}
\begin{center}
\leavevmode
\psfig{figure=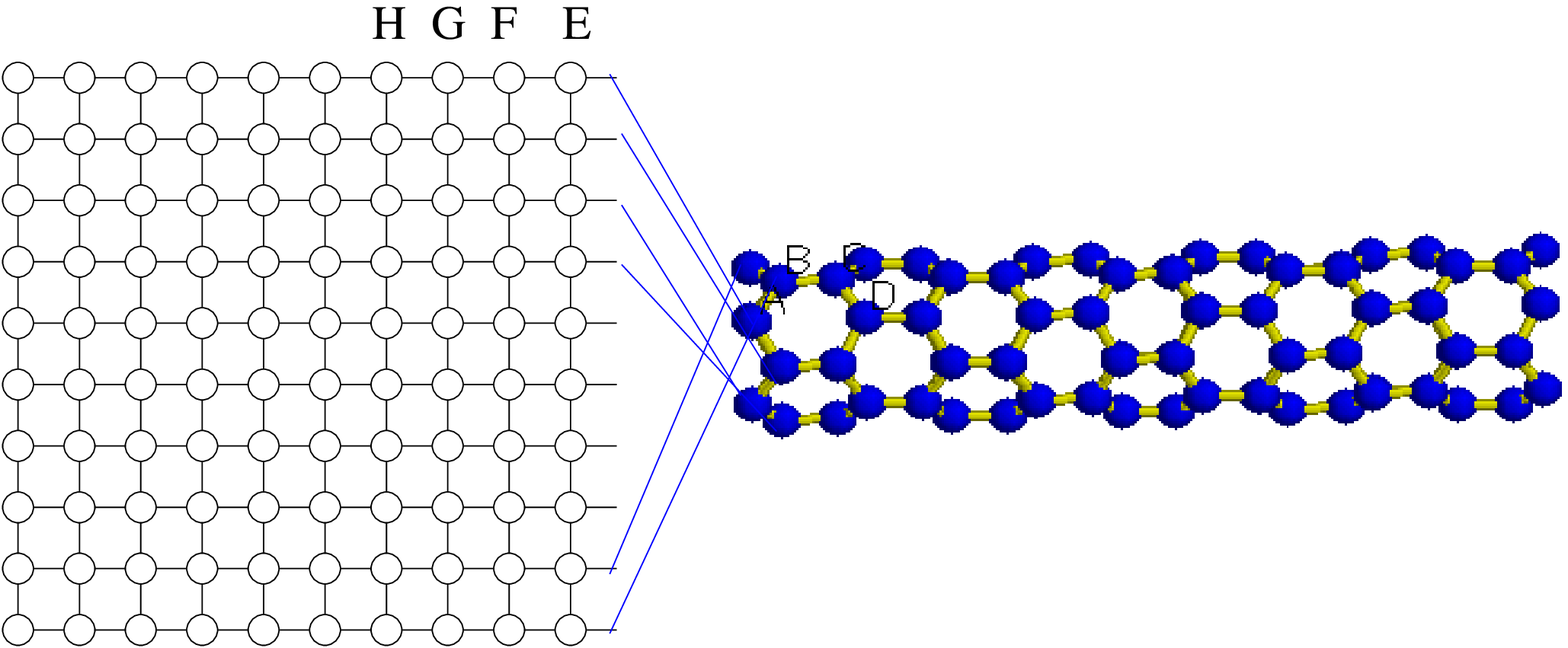,width=8cm,angle=0}
\end{center}
\caption{The
open zig-zag (5,0) nanotube composed of 10 layers 
end-connected to a two dimensional $10\times10$ superconductor.
}
\label{supra_swntzigzag.fig}
\end{figure}

\begin{figure}
\begin{center}
\leavevmode
\psfig{figure=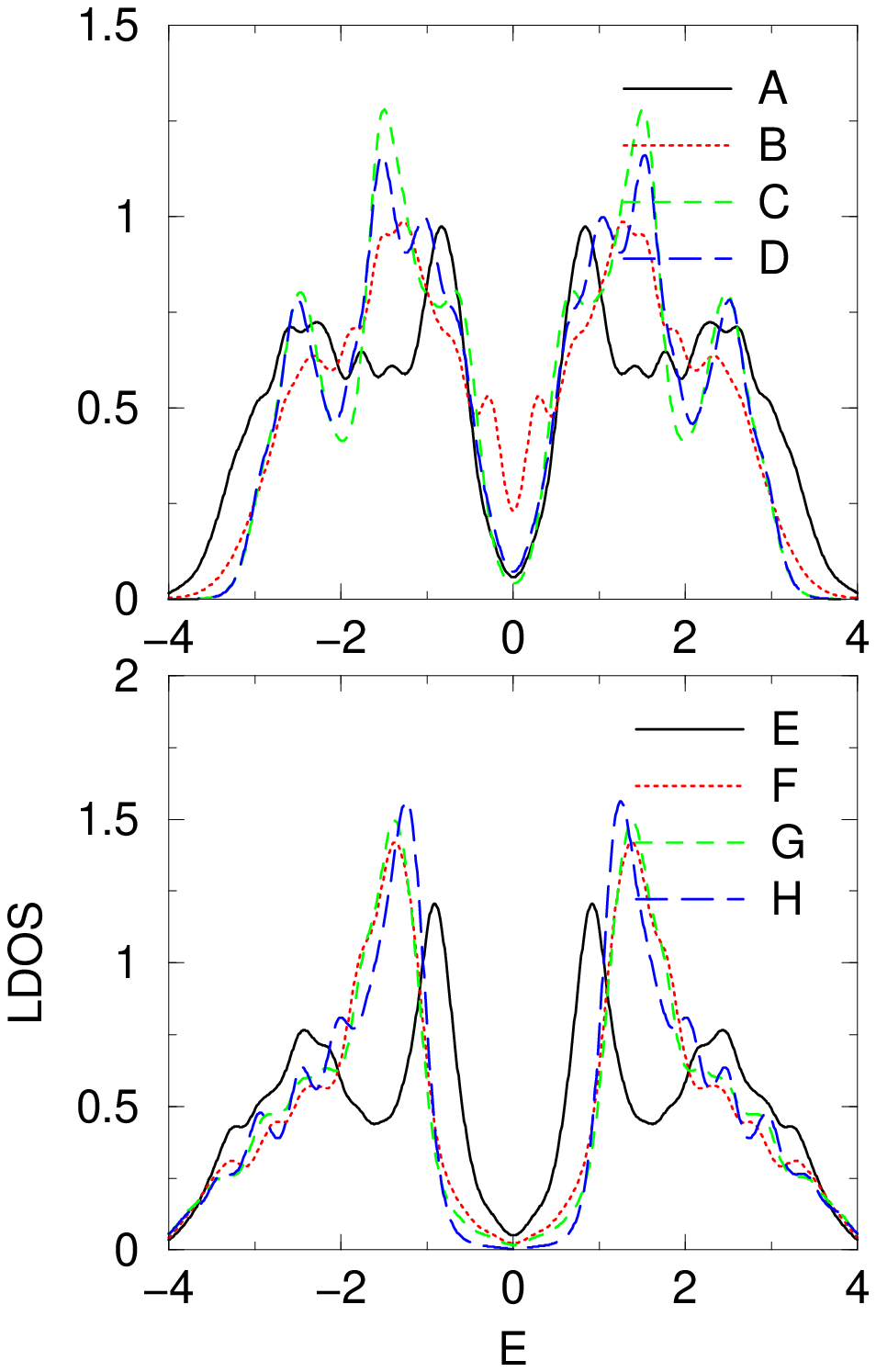,width=6cm,angle=0}
\end{center}
\caption{
The LDOS for the hybrid structure shown in Fig. \ref{supra_swntzigzag.fig}.
The points A,B,C,D belong to the end-connected nanotube
while the points E,F,G,H to the s-wave superconductor.
}
\label{LDOSsupra50.fig}
\end{figure}

We present now the proximity effect in superconductor - zig zag 
nanotube junction seen in Fig. \ref{supra_swntzigzag.fig}. 
We see that differently to the armchair case, inside nanotube 
the LDOS practically does not change with position 
(see \ref{LDOSsupra50.fig}). Also the LDOS is reduced due to 
the semiconducting character of the material. 
Inside superconductor we see that the LDOS recovers the bulk value 
in few lattice layers from the surface while in the armchair case 
the bulk value appears for larger distance. 

\begin{figure}
\begin{center}
\leavevmode
\psfig{figure=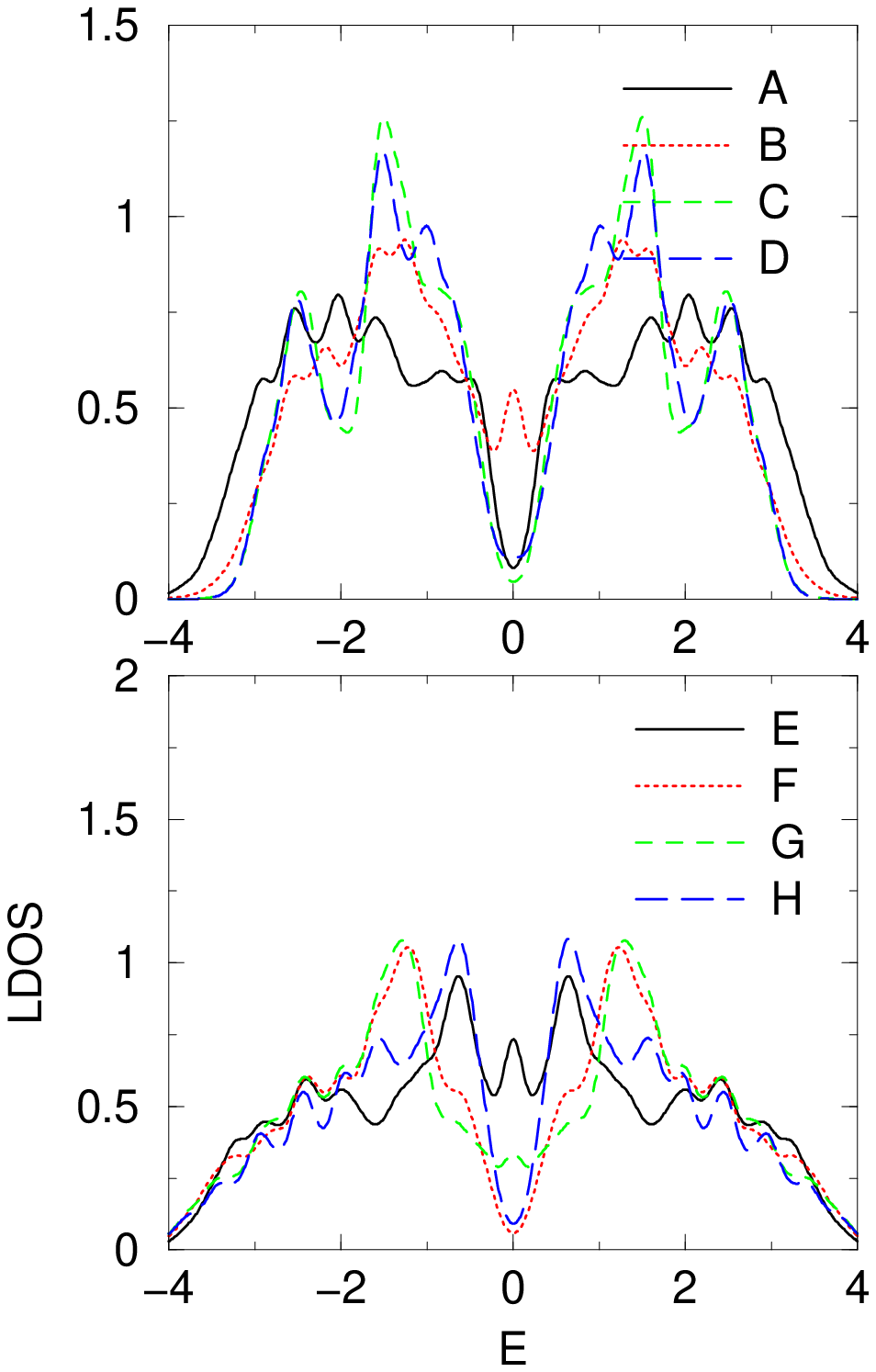,width=6cm,angle=0}
\end{center}
\caption{
The LDOS for the hybrid structure shown in Fig. \ref{supra_swntzigzag.fig}.
The points A,B,C,D belong to the end-connected nanotube
while the points E,F,G,H to the d-wave superconductor.
}
\label{LDOSdsupra50.fig}
\end{figure}

We also tested the case of the different pairing symmetry i.e. d-wave. 
As seen in Fig. \ref{LDOSdsupra50.fig} the main difference with 
the armchair case is the appearance of the ZBCP \cite{kashiwaya,alff}. 
This is attributed to 
the insulating character of nanotube which causes reflection of 
quasiparticles from the surface and appearance of peak due to 
the sign change of the pair amplitude.
We note that the interface is along the $[100]$ direction where 
for usual junctions the ZBCP is not expected. However in the 
present case the appearance of ZBCP is due to the distortion of the 
interface from the $[100]$ direction by the honeycomb lattice.
The conclusion from this section is that the proximity effect is 
reduced for superconductor zig-zag nanotube due to the 
insulating character of the nanotube. We can provide an 
additional explanation in terms of Andreev reflection 
which is responsible for the proximity effect. The Andreev 
reflection is modified for superconductor zig-zag nanotube 
due to the absence of charge carriers in the SWNT. 
As 
a consequence for d-wave superconductor zig-zag nanotube 
hybrid structure, ZBCP appears 
similarly to the appearance of ZBCP in d-wave superconductors having 
the appropriate orientation,
close to rigid insulating surfaces, where the reflected quasiparticles 
experience different sign of the pair amplitude.

\subsection{more realistic geometries}
\begin{figure}
\begin{center}
\leavevmode
\psfig{figure=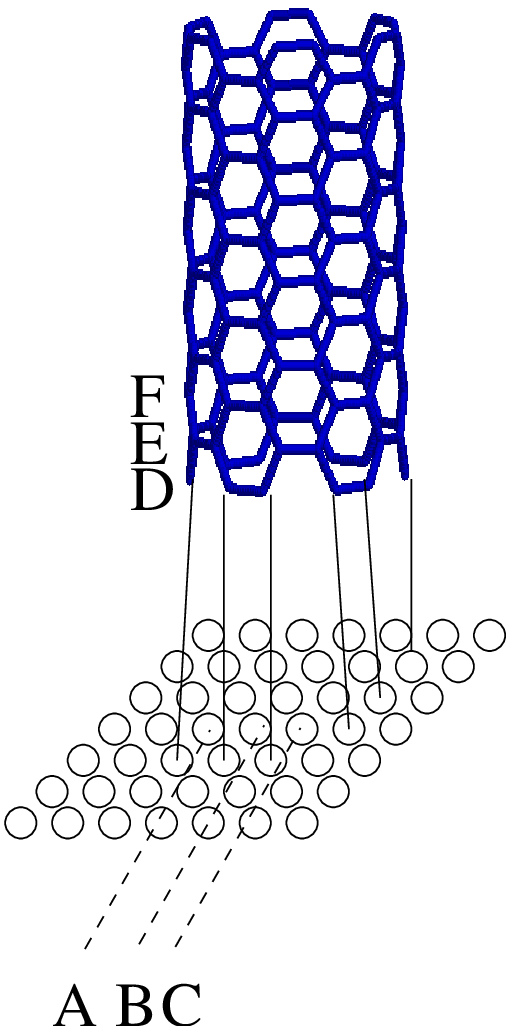,width=4cm,angle=0}
\end{center}
\caption{The
open zig-zag (6,6) nanotube composed of 12 layers 
top-connected to a two dimensional $12\times12$ superconductor.
}
\label{topsupra_swnt.fig}
\end{figure}

\begin{figure}
\begin{center}
\leavevmode
\psfig{figure=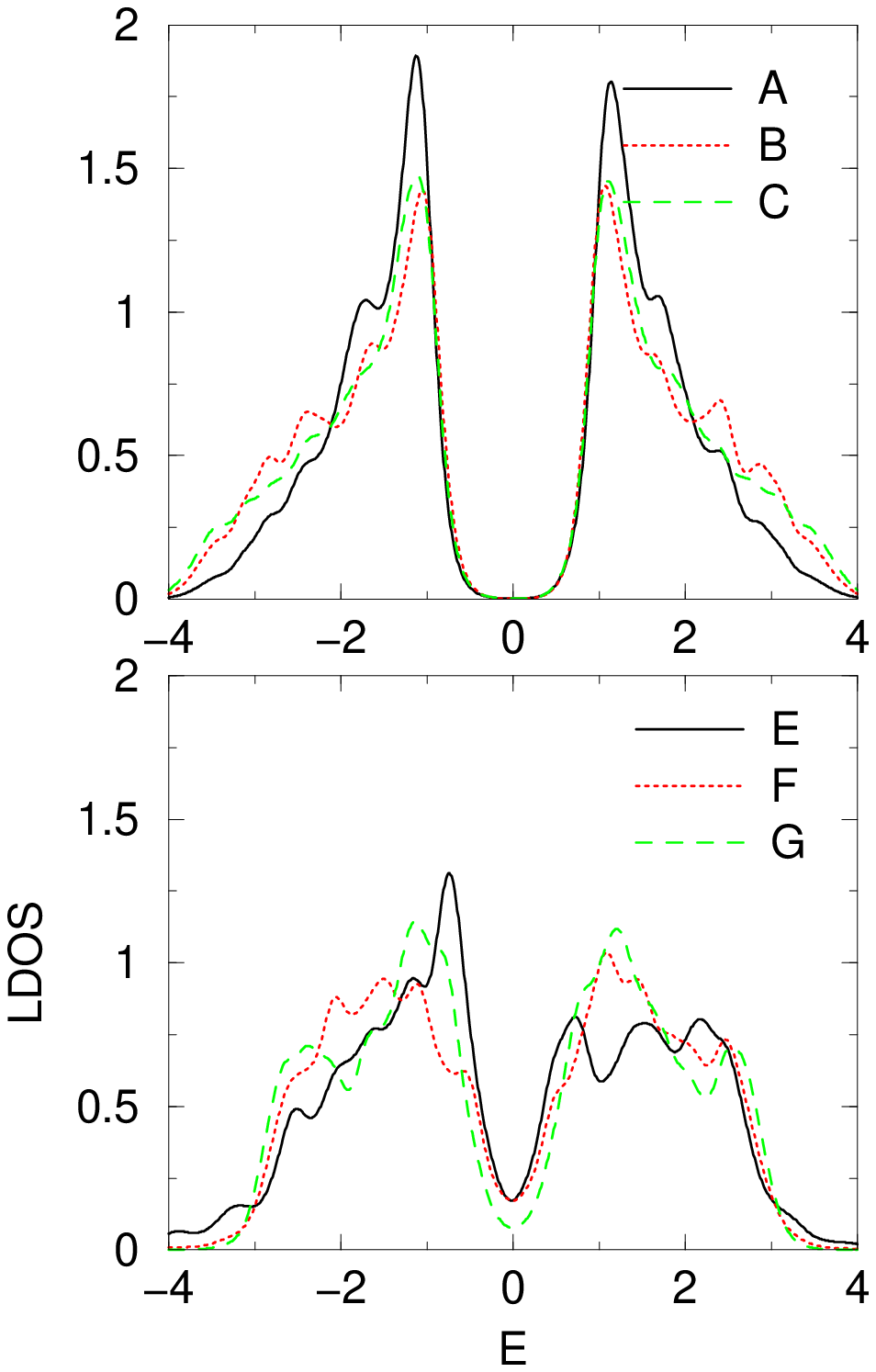,width=6cm,angle=0}
\end{center}
\caption{
The LDOS for the hybrid structure shown in Fig. \ref{topsupra_swnt.fig}.
The points A,B,C belong to the superconductor
while the points D,E,F to the top-connected nanotube.
}
\label{topLDOS66supra_swnt.fig}
\end{figure}

\begin{figure}
\begin{center}
\leavevmode
\psfig{figure=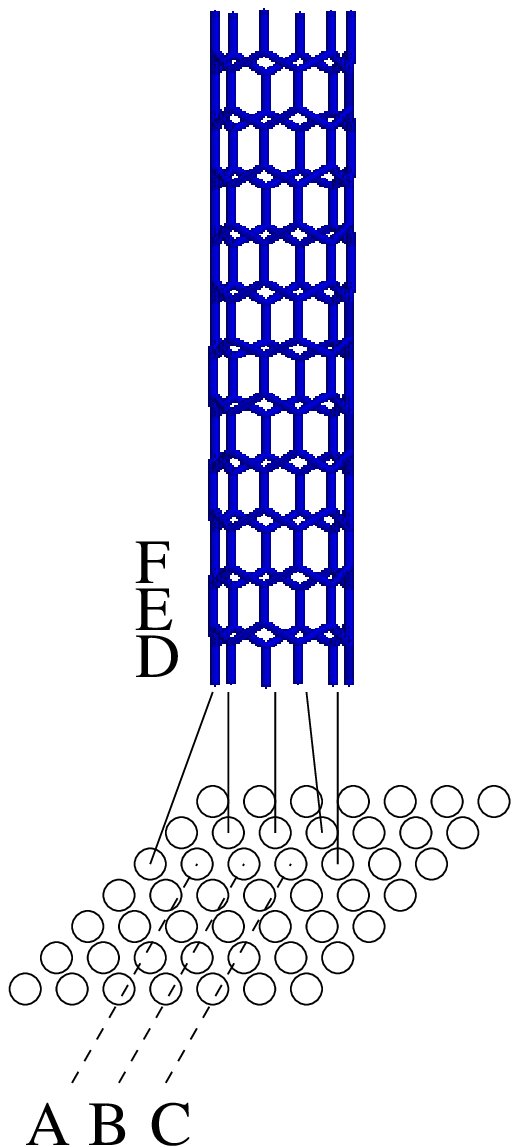,width=6cm,angle=0}
\end{center}
\caption{The
open zig-zag (6,0) nanotube composed of 12 layers 
top-connected to a two dimensional $12\times12$ superconductor.
}
\label{topsupra_swnt60.fig}
\end{figure}

\begin{figure}
\begin{center}
\leavevmode
\psfig{figure=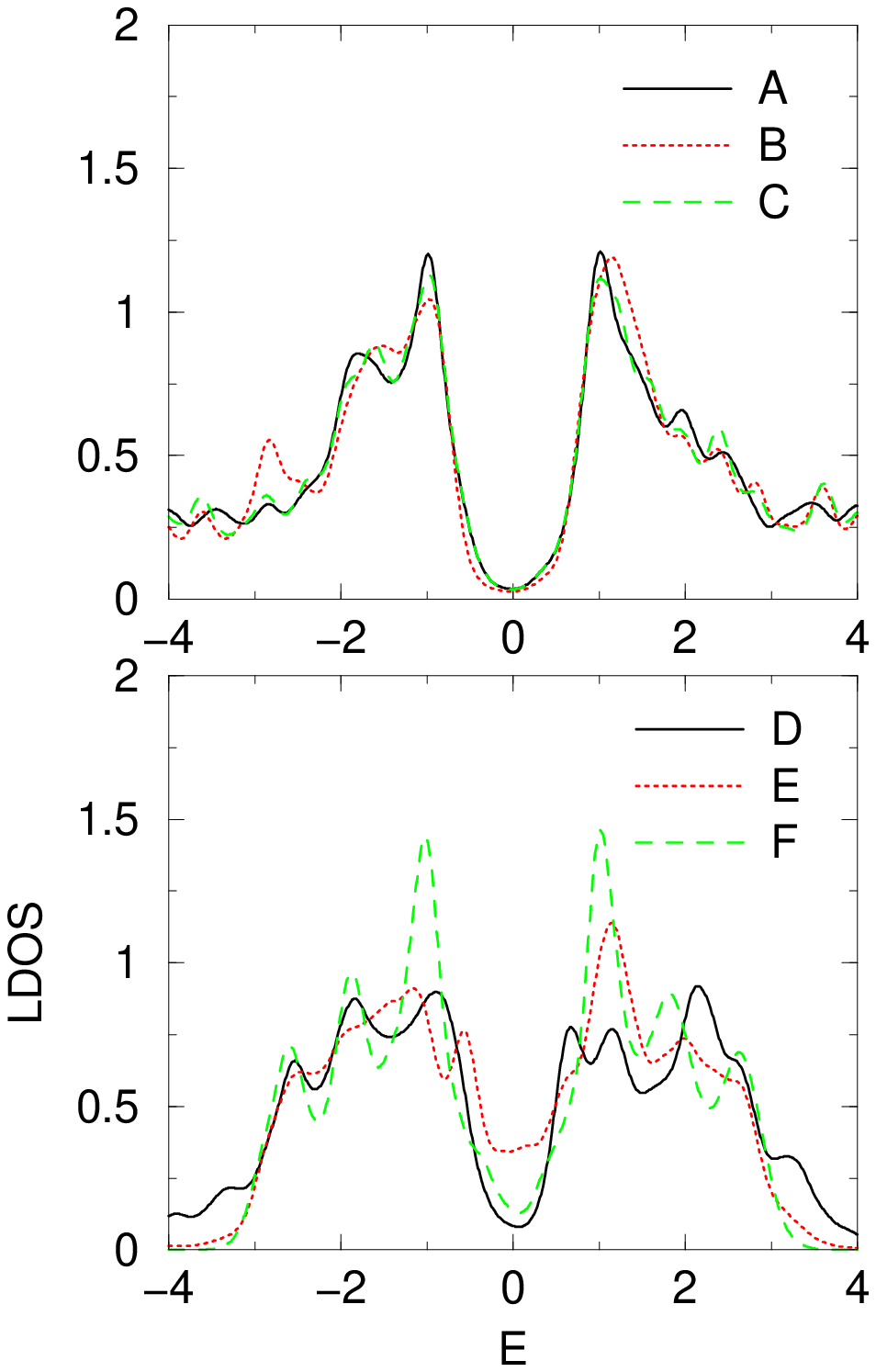,width=6cm,angle=0}
\end{center}
\caption{
The LDOS for the hybrid structure shown in Fig. \ref{topsupra_swnt60.fig}.
The points A,B,C belong to the superconductor
while the points D,E,F to the top-connected nanotube.
}
\label{topLDOS60supra_swnt.fig}
\end{figure}

Finally we would like to present the more realistic geometry
where the nanotube is connected on top of the two 
dimensional superconducting film as shown in 
Fig. \ref{topsupra_swnt.fig}. We restrict here to 
s-wave superconductors since as is well known d-wave 
superconductivity actually occurs in the 2-D plane and 
important effects due to sign change of the pair 
amplitude can not be observed when tunneling in the 
z-direction is considered. 
The LDOS presented for open armchair nanotube 
in Fig. \ref{topLDOS66supra_swnt.fig} shows 
that no important conclusions can be drawn compared to the 
previous connection geometries. Also here the gap coexists 
with the one dimensional bands. However the present structure 
has the advantage that novel phenomena may appear due to the 
confinement of the superconducting sites inside the restricted 
region which is defined by the periphery of the tube. This 
may become more obvious when the transparency of the interface 
is low and the superconducting atoms are encapsulated inside the 
tube. In the latter case we may speak about few atom 
nano grain superconductivity. 
In order to test the latter case more precisely we present the 
case where the superconductor is connected to a zigzag 
nanotube as shown in Fig. \ref{topsupra_swnt60.fig}. 
Here the interface is of reduced transparency due to the 
insulating features of the zig-zag nanotube.
The LDOS  ( see Fig. \ref{topLDOS60supra_swnt.fig}) shows 
residual states inside the gap due to the semiconducting 
$(6,0)$ nanotube which acts a pair breaking mechanism. 

\section{conclusions}
We studied the electronic properties of SWNT - superconductor 
hybrid structures with in the 
Hubbard model self consistently. The results indicate that the 
LDOS is strongly modified close to the boundary layers of the 
nanotube as well as close to the interface of junctions 
with superconducting materials.
We showed
that the proximity LDOS between the superconductor 
and the nanotube
can be modulated by simple geometrical transformations like
rotation of the SWNT with respect to the superconductor. 
We found that the proximity induced gap is reduced for 
side connection. 
We demonstrated that the proximity effect depends on the chirality 
of the nanotube. We provided the explanation in terms of modified
Andreev reflection in front of a metallic (armchair) or 
insulating (zigzag) interfaces. So one could say that the 
proximity effect can be viewed as a novel way to classify the nanotubes 
in metallic or semiconducting ones.
Finally we found that the LDOS is sensitive to the pairing symmetry 
of the superconductor and shows features due to the 
geometric structure of the nanotube.
In the last section we showed that the results that we presented here 
can be extended to more realistic geometries, and nanotubes 
with larger diameters where hybridization effects are 
absent and the approach what we used here is valid.

\section{acknowledgments}
Part of this work was done in University of Tuebingen. 
 
\bibliographystyle{prsty}


\vspace{3cm}
\begin{figure}
\begin{center}
\leavevmode
\psfig{figure=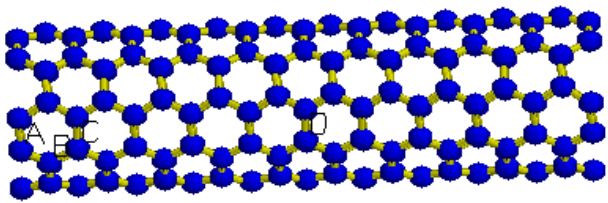,width=10cm,angle=0}
\end{center}
\caption{The open armchair (5,5) nanotube composed of 21 layers.
}
\label{55.fig}
\end{figure}

\begin{figure}
\begin{center}
\leavevmode
\psfig{figure=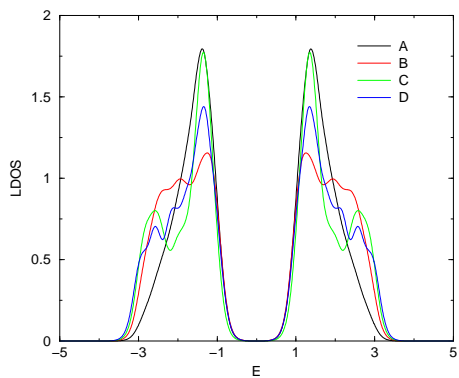,width=6cm,angle=0}
\end{center}
\caption{The LDOS for points A,B,C,D of an armchair (5,5) SWNT 
shown in Fig. \ref{55.fig} exhibiting superconductivity.
}
\label{LDOS55supra.fig}
\end{figure}

\end{document}